\documentclass[twocolumn,prl,aps,showpacs,preor==]{revtex4-1}
\pdfoutput=1
\usepackage{preprintcover}
\usepackage{graphicx}
\usepackage{dcolumn}
\usepackage{multirow}
\usepackage{amssymb}
\usepackage{amsmath}
\usepackage{slashed}
\usepackage{atlasphysics}
\usepackage{hyperref}
\usepackage{pifont}
\usepackage[usenames,dvipsnames]{xcolor}

\newcommand{\PbPb}{\mbox{Pb+Pb}}
\newcommand{\pp}{\mbox{$pp$}}

\newcommand{\ANpart}{\mbox{$\langle N_{\mathrm{part}}\rangle$}}

\newcommand{\Njet}{\mbox{$N_{\mathrm{jet}}$}}

\newcommand{\sqrtsnn}{\mbox{$\sqrt{s_{\mathrm{NN}}}$}}

\newcommand{\RAA}{\mbox{$R_{\rm AA}$}}

\newcommand{\RFour}{\mbox{$R = 0.4$}}

\newcommand{\RTwo}{\mbox{$R= 0.2$}}

\newcommand{\ETfcal}{\mbox{$\Sigma E_{\mathrm{T}}^{\mathrm{FCal}}$}}

\newcommand{\invnb}{\mbox{${\rm nb^{-1}}$}}
\newcommand{\kt}{\mbox{$k_{t}$}}

\newcommand{\Dphi}{\mbox{$\Delta \phi$}}

\newcommand{\pTjet}{\mbox{$p_{{\mathrm{T}}}$}}

\newcommand{\pTtrkjet}{\mbox{$p_{{\mathrm{T}}}^{\mathrm{trkjet}}$}}

\newcommand{\vtcalo}{\mbox{$v_2^{\mathrm{calo}}$}}

\newcommand{\vtjet}{\mbox{$v_2^{\mathrm{jet}}$}}
\newcommand{\vtjetmeas}{\mbox{${v_2^{\mathrm{jet}}|_{\mathrm{meas}}}$}}
\newcommand{\vtwo}{\mbox{$v_2$}}

\newcommand{\Psit}{\mbox{$\Psi_2$}}

\newcommand{\Psires}{\mbox{$\kappa$}}

\newcommand{\Rmaxdphi}{\mbox{$R^{\mathrm{max}}_{\Delta \phi}$}}
\newcommand{\diff}{\mbox{$d$}}
\newcommand{\dphi}{\mbox{$\Delta\phi$}}
\newcommand{\dNdpTdphi}{\mbox{$\diff^2\Njet/\diff\pt\diff\dphi$}}
\newcommand{\dNdpTdphiMeas}{\mbox{$\dfrac{\diff^2N_{\mathrm{jet}}^{\mathrm{meas}}}{\diff\pt\diff\dphi}$}}
\newcommand{\dNdpTdphiCorrTall}{\mbox{$\dfrac{\diff^2N_{\mathrm{jet}}^{\mathrm{corr}}}{\diff\pt\diff\dphi}$}}

\newcommand{\ftwo}{\mbox{$f_2$}}
\newcommand{\ftwocorr}{\mbox{$f_2^{\mathrm{corr}}$}}

\newcommand{\emdash}{--}
\newcommand{\outplane}{\mbox{$3\pi/8 \leq \Dphi \leq \pi/2$}}
\newcommand{\inplane}{\mbox{$0 \leq \Dphi < \pi/8$}}

\newcommand{\opencircle}{\mbox{$\,\circ\,$}}

\newcommand{\fullcircle}{\mbox{$\,\bullet\,$}} 
\newcommand{\fullsquare}{\,\vrule height5pt depth0pt width5pt\,}
\definecolor{kAzure}{rgb}{0.0,0.4,0.8}
\newcommand{\PiOverTwo}{\mbox{$\pi/2$}}

\newcommand{\papertype}{Letter}
\newcommand{\papertitle}{Measurement of the Azimuthal Angle Dependence of
    Inclusive Jet Yields in \PbPb\ Collisions at $\sqrtsnn = 2.76$~\TeV\ with the ATLAS detector}

\newcommand{\paperabstract}{
Measurements of the variation of inclusive jet suppression as a
function of relative azimuthal angle, \Dphi, with respect to the elliptic event
plane provide insight into the path-length dependence of jet
quenching. ATLAS has measured the \Dphi\ dependence of jet
yields in 0.14~$\mathrm{nb}^{-1}$ of $\sqrtsnn = 2.76$~\TeV\ \PbPb\ collisions at the LHC for 
jet transverse momenta $\pT > 45$~\GeV\ in different
collision centrality bins using an underlying event subtraction procedure that accounts for elliptic
flow. The variation of the jet yield with \Dphi\ was characterized by
the parameter, \vtjet, and the ratio of 
out-of-plane ($\Dphi \sim \PiOverTwo$) to in-plane ($\Dphi\sim 0$)
yields. Non-zero \vtjet\ values were measured in all   
centrality bins for $\pT < 160$~\GeV. The jet yields are observed to
vary by as much as 20\% between in-plane and out-of-plane directions.}
\PreprintIdNumber{CERN-PH-EP-2013-081}
\PreprintCoverPaperTitle{\papertitle}
\PreprintCoverAbstract{\paperabstract}
\PreprintJournalName{PRL}

\begin{document}
 \preprint{APS/123-QED}

\title{\papertitle}
\author{ATLAS Collaboration}
\begin{abstract}
\paperabstract
\end{abstract}
\maketitle 

Studies of jet production in \PbPb\ collisions at the LHC
\cite{Aad:2010bu,Chatrchyan:2011sx} show behavior consistent with
``jet quenching,'' a general term for the modification of parton
showers in the hot dense medium created in ultra-relativistic nuclear collisions. For
example, the inclusive yield of jets was observed to be suppressed by
a factor of approximately two in central \PbPb\ collisions relative to
peripheral collisions \cite{jetsupp}, consistent with a medium-induced
reduction in the jet energies. Perturbative or weak-coupling
calculations model jet energy loss, $dE/dx$,  through a combination of
collisional and radiative energy loss of the constituents of the
parton shower. The radiative contributions are subject to coherence
effects \cite{Baier:1998yf} that explicitly depend on the path length
of the shower in the medium. Strong-coupling calculations suggest a
different path-length dependence \cite{Gubser:2008as,Marquet:2009eq}.
Measurements of the jet yield as a function of quantities providing
indirect control over the jet path lengths  may provide insight into
the physical mechanisms responsible for jet quenching
\cite{Gyulassy:2000gk,Zhang:2012ha}.  Such quantities include the
\PbPb\ collision centrality and the azimuthal angle of the jet with
respect to the elliptic event plane.

Elliptic flow refers to a $\cos{2\phi}$ modulation of the azimuthal
angle ($\phi$) distribution of particles produced in
ultra-relativistic nuclear collisions~\footnote{ATLAS uses a
  right-handed coordinate system with its origin at the nominal
  interaction point (IP) in the center of the detector and the
  $z$-axis along the beam pipe. The $x$-axis points from the IP to the
  center of the LHC ring, and the $y$-axis points upward. Cylindrical
  coordinates $(r,\phi)$ are used in the transverse plane, $\phi$
  being the azimuthal angle around the beam pipe. The pseudorapidity
  is defined in terms of the polar angle $\theta$ as
  $\eta=-\ln\tan(\theta/2)$.}. This modulation is understood to arise
from an approximately elliptic anisotropy of the initial-state
transverse energy density  profile that is imprinted on the azimuthal
angle distribution of final-state particles \cite{Ollitrault:1992bk}
by the strong collective evolution of the medium. The resulting
azimuthal angle distribution is often parameterized by the form:
\begin{equation} \frac{dN}{d\phi} \propto 1 + 2\vtwo \cos 2(\phi - \Psi_2),
\label{eq:ellipticmod}
\end{equation}
where the elliptic event plane angle, $\Psi_2$, specifies the
orientation of the initial density profile in the transverse plane,
and the parameter \vtwo\ quantifies the magnitude of the
modulation. Jets measured at different azimuthal angles relative to
the event plane, $\Dphi \equiv \phi_{\mathrm{jet}} - \Psi_2$, result
from parton showers that traverse, on average, different path lengths
and density profiles in the medium. Thus, a measurement of the
variation of the jet yield as a function of \Dphi\ should provide a
direct constraint on theoretical models of the path-length dependence
of the energy loss. This measurement is not directly sensitive to the
contribution to jet suppression from higher order flow harmonics,
which may arise from the fluctuating initial geometry
\cite{ALICE:2011ab,ATLAS:2012at,Chatrchyan:2012wg}.

Variations in jet yield as a function of \Dphi\ have been observed
indirectly through measurements of single hadrons with large transverse
momentum (\pT) at RHIC \cite{Adams:2004wz,Adler:2006bw,Adare:2010sp}
and the LHC \cite{ATLAS:2011ah, Chatrchyan:2012xq, Abelev:2012di}. The
utility of such measurements is limited by the weak relationship
between hadron \pt\ and the transverse momentum of the parent parton
shower. This \papertype\ presents results of  measurements using fully
reconstructed jets, which have kinematics properties that are more closely
related to those of the parent partons. The \Dphi\ dependence of the
inclusive jet yield was measured in $\sqrtsnn = 2.76$~\TeV\ \PbPb\
collisions as a function of jet \pT\ and \PbPb\ collision
centrality. The measurement was performed with the anti-\kt\ algorithm
\cite{Cacciari:2008gp} with distance parameter \RTwo, chosen to limit
the contribution of the underlying event (UE) to the measurement. The
\Dphi\ dependence was characterized by the jet \vtwo, \vtjet, and the
ratios of the jet yields in different \Dphi\ intervals at fixed \pt\
and centrality. Such dependence is expected to be small in either the
most central or most peripheral collisions, due to the lack of
initial-state anisotropy and the lack of quenching, respectively.  For
intermediate centralities, measurement of the \Dphi\ dependence of the
jet yields probes the interplay between contributions to quenching
from the overall system size and energy density, as well as from the
initial state anisotropy.

The measurements presented here were performed with the ATLAS detector
\cite{Aad:2008zzm} using its calorimeter, inner detector, trigger, and
data acquisition systems. The calorimeter system consists of a liquid
argon (LAr) electromagnetic (EM) calorimeter covering $|\eta|<3.2$, a
steel/scintillator sampling hadronic calorimeter covering $|\eta| <
1.7$, a LAr hadronic calorimeter covering $1.5 < |\eta| < 3.2$, and a
forward calorimeter (FCal) covering $3.2 < |\eta| <
4.9$. Charged-particle tracks were measured over the range
$|\eta|<2.5$ using the inner detector \cite{Aad:2010bx}, which is
composed of silicon pixel detectors in the innermost layers, followed
by silicon microstrip detectors and a straw-tube transition-radiation
tracker ($|\eta| < 2.0$), all immersed in a 2~T axial magnetic
field. The zero-degree calorimeters (ZDCs) are located
symmetrically at $z = \pm 140$~m and cover $|\eta| > 8.3$. In
\PbPb\ collisions the ZDCs primarily measure non-interacting ``spectator''
neutrons from the incident nuclei. A ZDC coincidence trigger was
defined by requiring a signal consistent with one or more neutrons in
each of the calorimeters.

\PbPb\ collisions corresponding to a total integrated luminosity of
$0.14~\invnb$ were analyzed. The events were recorded using either a
minimum-bias trigger, formed from the logical OR of triggers based on
a ZDC coincidence or total transverse energy in the event, or a jet
trigger implemented using the \PbPb\ jet reconstruction algorithm. The
jet trigger selected events having at least one jet with transverse
energy $\ET > 20$~\GeV. Event selection and background rejection
criteria were applied \cite{ATLAS:2011ah} yielding 52 million and 14
million events in the minimum-bias and jet-triggered samples,
respectively. For each event, $\Psi_2$ was computed from the azimuthal
distribution of the transverse energy measured in the FCal
\cite{Poskanzer:1998yz,ATLAS:2011ah}, and angles with respect to
$\Psi_2$ were defined over $0 \leq \Dphi \leq \PiOverTwo$. The
centrality of \PbPb\ collisions was characterized by \ETfcal, the
total transverse energy measured in the FCal \cite{ATLAS:2011ah}. The
results reported here were obtained using the following centrality
intervals defined according to successive percentiles of the \ETfcal\
distribution ordered from the most central (highest \ETfcal) to the
most peripheral collisions: 5\emdash10\%, 10\emdash20\%,
20\emdash30\%, 30\emdash40\%, 40\emdash50\%, and 50\emdash60\%. The
centrality interval 5--60\% coincides to the range over which the
$\Psi_2$ resolution is adequate for the measurement. The percentiles
were defined after correcting the \ETfcal\ distribution for an
estimated 2\% minimum-bias trigger inefficiency that is concentrated
in the most peripheral intervals, which are not included in this
analysis.  A Glauber model analysis \cite{Miller:2007ri,Alver:2008aq}
of the \ETfcal\ distribution \cite{ATLAS:2011ah} was used to evaluate
the average number of nucleons participating in the collision,
\ANpart, in each centrality interval.

The jet reconstruction and underlying event subtraction procedures are
the same as those used in Ref.~\cite{jetsupp}, which is summarized in
the following. The anti-\kt\ algorithm was applied to calorimeter towers
with segmentation $\Delta\eta\times\Delta\phi=0.1\times0.1$. A
two-step iterative procedure was used to obtain an event-by-event
estimate of the average \eta-dependent UE energy density while
excluding actual jets from that estimate. The jet kinematics were
obtained by subtracting the UE energy from the towers within the
jet. This subtraction accounts for elliptic flow by modulating the
average background density by the magnitude of the elliptic flow
measured by the calorimeter, \vtcalo, over the interval $|\eta|<3.2$
and excluding $\eta$ regions containing jets. Following
reconstruction, the jet energies were corrected to account for the
calorimeter energy response using an $\eta$- and \ET-dependent
multiplicative factor that was derived from the MC simulation \cite{Aad:2011he}.

Separate from the calorimeter jets, ``track jets'' were reconstructed
by applying the anti-\kt\ algorithm with \RFour\ to charged particles
having $\pt > 4$~\GeV. The \pT\ of the track jets, \pTtrkjet, is
largely unaffected by the UE due to the $\pt > 4$~\GeV\
requirement. To exclude the contribution to the jet yield from UE
fluctuations of soft particles falsely identified as calorimeter jets, the jets
used in this analysis were required to be within $\Delta R
=\sqrt{\Delta\eta^2+\Dphi^2} =0.2$ of a track jet with $\pTtrkjet>
10$~\GeV\ or an EM cluster \cite{Aad:2009wy} with $\pt > 9$~\GeV\
\footnote{This differs from the requirement used in
Ref.~\cite{jetsupp}, where the minimum \pt\ was 7~\GeV\ for both track
jets and clusters.}.

The performance of the jet reconstruction was evaluated using the
GEANT4-simulated detector response \cite{Agostinelli:2002hh,atlassim}
in a Monte Carlo (MC) sample of \pp\ hard scattering events at
$\sqrtsnn = 2.76$~\TeV. The events were produced with the PYTHIA event generator 
\cite{Sjostrand:2006za} version 6.423 using the AUET2B tune
\cite{mctunes} and overlaid on minimum-bias \PbPb\ collisions
recorded by ATLAS. Through this embedding procedure, the MC sample
contains a UE contribution that is identical in all respects to the
data, including azimuthal modulation of the UE due to harmonic
flow. Jets reconstructed in the MC events using the same algorithms as
applied to data were compared to generator-level jets reconstructed
from final-state PYTHIA hadrons. Potential variations in the jet
energy resolution (JER) and jet energy scale (JES) with \Dphi\ due to
elliptic and higher-order modulation
\cite{ALICE:2011ab,Chatrchyan:2012wg,ATLAS:2012at} of the UE were
investigated in  the MC sample; no significant variation was found. 

The dependence of the JES on \Dphi\ was further constrained by
comparing the calorimeter jets to matched track jets in the data. For different
values of \Dphi, the mean \pt\ of calorimeter jets was evaluated as a
function of the \pt\ of the matched track jet, and no significant
variation with \Dphi\ was observed. This study provides an upper limit
on the variation in the JES between jets at $\Dphi=0$ and
$\Dphi=\PiOverTwo$ of 0.1\% for $\pTjet>45$~\GeV.
\begin{figure}
\centerline{ \includegraphics[width=0.45
  \textwidth]{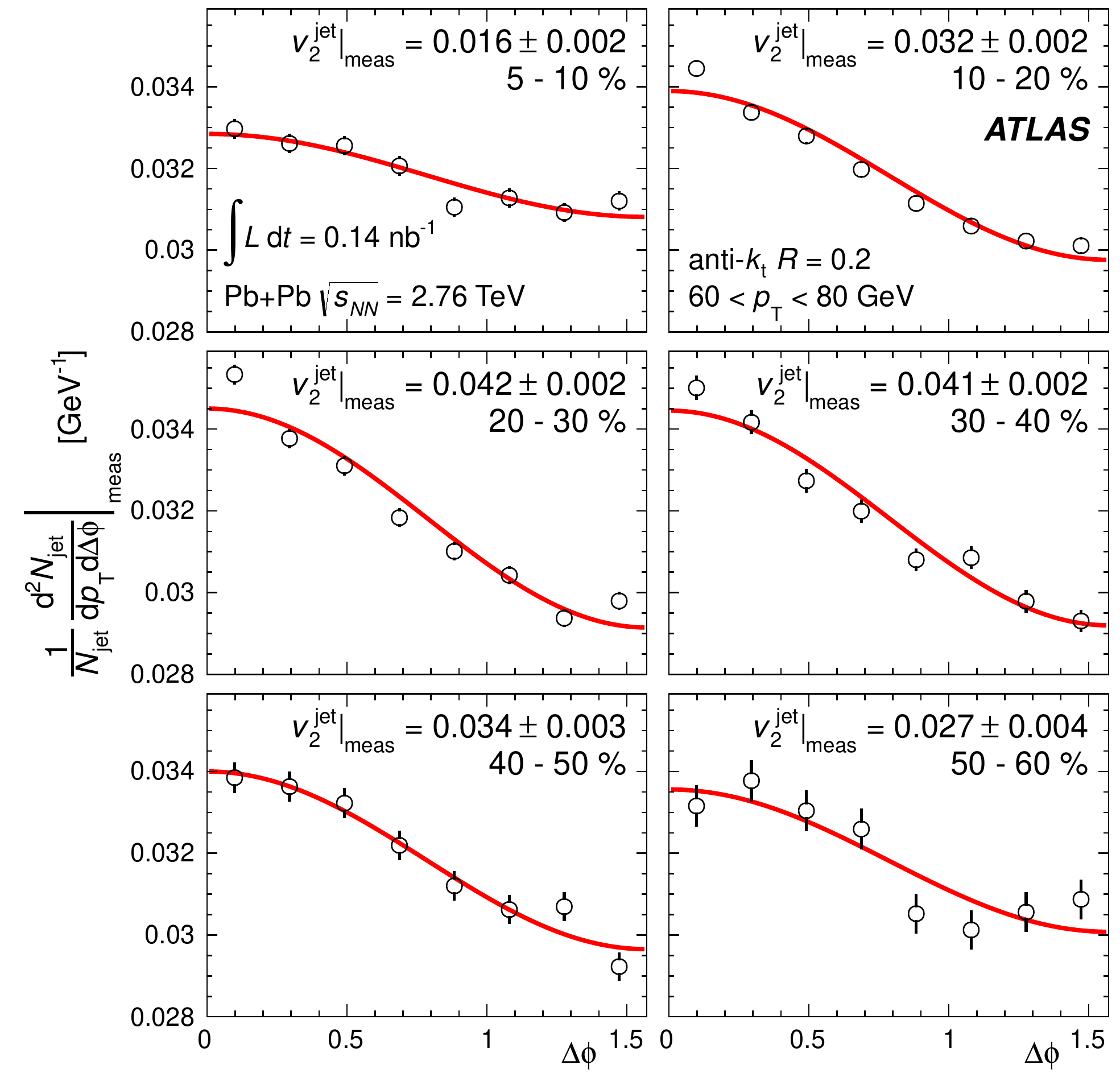} } \caption{ \Dphi\
  dependence of measured \dNdpTdphi\ in the $60 < \pt < 80$~\GeV\
  interval for six ranges of collision centrality. The yields are
  normalized by the total number of jets in the \pt\ interval. The
  solid curves indicate the results of  fitting the data to the
  functional form of Eq.~(\ref{eq:ellipticmod}), with the resulting
  \vtwo\ values, \vtjetmeas, listed in each panel. The error bars and
  errors on \vtjetmeas\ indicate statistical uncertainties. }
\label{fig:uncorrmod}
\end{figure}

Double differential  jet yields, \dNdpTdphi, were measured over
$|\eta| < 2.1$ for each of the centrality ranges described above and
in five \pTjet\ intervals:  45\emdash60 \GeV, 60\emdash80 \GeV,
80\emdash110 GeV, 110\emdash160 \GeV, and 160\emdash210 \GeV. The
measurement in each \pTjet\ range was performed using events selected
by the jet trigger except for the 45\emdash60~\GeV\ \pT\ range, in
which minimum-bias events were used. The \Dphi\
dependence of the jet yields in the 60\emdash80~\GeV\ \pTjet\ interval
is shown for each centrality range in Fig.~\ref{fig:uncorrmod}. A
significant \Dphi\ variation that is consistent with a $\cos{2\Dphi}$
modulation is seen for all centrality intervals.

The measured yields and the resulting \vtjetmeas\ values are distorted
by the finite resolutions in $\Psi_2$ and the jet \pT. The $\Psi_2$
resolution was evaluated using a sub-event technique
\cite{Poskanzer:1998yz,ATLAS:2011ah} in which \Psit\ was measured
separately in the positive and negative $\eta$ halves of the FCal
yielding values $\Psi_2^{+}$ and $\Psi_2^{-}$, respectively. The width
of the $\Psi_2^{+}-\Psi_2^{-}$ distribution was used
\cite{Poskanzer:1998yz} to estimate a factor, \Psires, that was used
to correct each measured \vtwo\ value for the finite \Psit\ resolution
according to  
\begin{equation} v_2 = v_2 |_{\mathrm{meas}}/\Psires\,. 
\label{eq:rescorr}
\end{equation}
This factor was evaluated for events containing jets to account for
the relevant distribution of events within each centrality
interval. 

The \pT\ dependence, and possibly also the \Dphi\ dependence, of the
measured yields are affected by the JER, which arises from both
fluctuations in the UE and the detector response. The MC study shows
that for the \RTwo\ jets used in this analysis, the JER-induced
migration between jet \pT\ intervals is sufficiently small that a
``bin-by-bin'' unfolding method, utilizing multiplicative corrections
to the jet yields, is appropriate. The bin-by-bin correction factors
are defined to be the number of generator-level jets divided by the
number of reconstructed jets in each \pTjet, \Dphi, and centrality
interval. The MC studies show no significant \Dphi\ variation of the
JER, JES, and the correction factors, and so these correction factors were
taken to be \Dphi-independent. Since the measurements presented here
depend only on the ratios of jet yields between different \Dphi\
intervals for the same \pt\ range, the correction factors do not
affect any of the final results; the potential for a \Dphi\
dependence of the correction factors is included in the estimates of
the systematic uncertainty.

Systematic uncertainties on the corrected \vtwo\ values arise due to
uncertainties on the two correction procedures described
above. Uncertainties on \Psires\ were estimated by using the values
obtained in previous studies\cite{ATLAS:2011ah} for slightly different
centrality intervals and interpolating to obtain values appropriate for
the centrality intervals used here. The uncertainties were found to vary between 1\% and 4\%
from central to peripheral collisions. Potential distortions in the
measurement of \Psit\ due to the production of jets in the FCal
pseudorapidity range were studied in the MC sample and found to be
negligible for the centrality intervals included in this analysis.

Uncertainties on the measurements arising from \Dphi-dependent
systematic uncertainties on the bin-by-bin correction factors were
estimated by determining the sensitivity of these correction factors
to each systematic variation and then parameterizing that sensitivity
with a $\cos2\Dphi$ dependence. The sensitivity to the \Dphi\
dependence of the spectrum was evaluated by varying the \pt\ spectrum
of the generator-level jets in each \Dphi\ interval within a range
consistent with the measured \vtjet\ values. The JES and JER
contributions to the uncertainty were obtained by varying the
relationship between generator-level and reconstructed jet \pt\ in the
determination of the correction factors. These procedures utilized the
JES constraints obtained from track jets and direct measurements of
the UE contribution to the JER \cite{jetsupp}. Parameterizations of
the measured \vtcalo\ and the average background \et\ underlying a
typical jet measured in the data were used to provide the dependence
of variations on centrality. The systematic uncertainties
on \vtjet\ are less than 0.002 except in the most central and lowest
\pt\ intervals where the JER contribution dominates, resulting in
uncertainties between 0.005 and 0.01.

\begin{figure}
\centerline{ \includegraphics[width=0.49\textwidth]{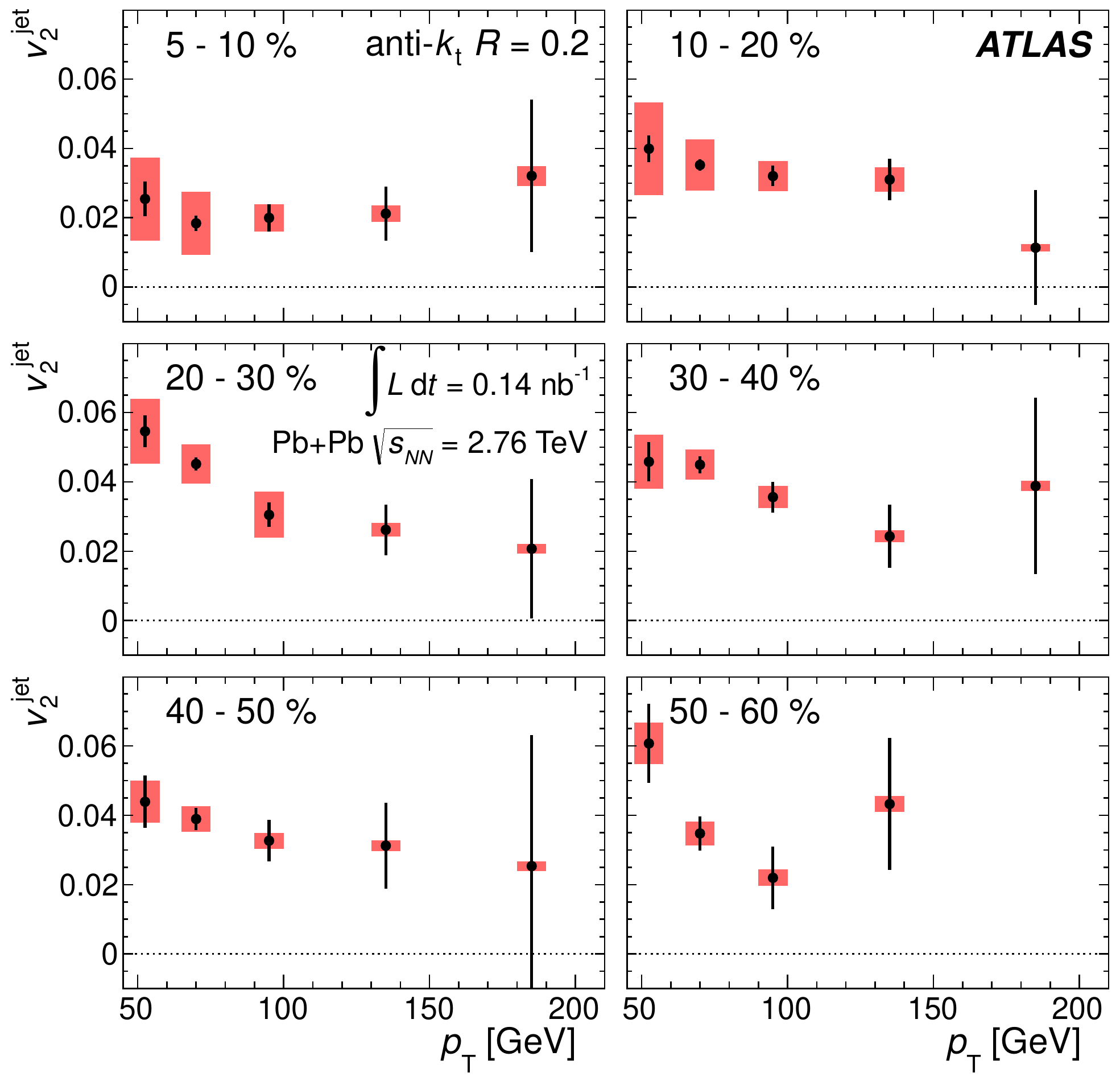}
}
 \caption{ \vtjet\ as a function of jet \pTjet\ in each centrality
   interval. The error bars on the points indicate statistical
   uncertainties while the shaded boxes represent the systematic
   uncertainties (see text). The horizontal width of the systematic
   error band is chosen for presentation purposes only. } 
\label{fig:jetvtwopt} 
\end{figure}

The azimuthal dependence of jet suppression can be characterized by
\vtjet, which was obtained by correcting the \vtjetmeas\  values using
Eq.(\ref{eq:rescorr}). Figure~\ref{fig:jetvtwopt} shows the resulting
\vtjet\ values as a function of jet \pTjet\ for all centrality
intervals. Significant, non-zero values are observed over the range
$45 < \pTjet < 160$~\GeV\ for all centrality intervals.
 A direct comparison between the \vtwo\ of single high-\pt\ charged
 particles and \vtjet\ is generally not possible; however the fact
 that both quantities exhibit only a weak \pt\ dependence leads to the
 expectation that they should be of similar magnitude. In the charged
 particle  measurements, the \vtwo\ values of charged particles with $28 < \pt
 < 48$~\GeV\ were found to vary between 0.02 and 0.05 for the
 10\emdash50\% centrality range \cite{Chatrchyan:2012xq}, which are
 generally in agreement with \vtjet\ values reported here indicating
 no obvious inconsistencies between the two results.

\begin{figure}
 \centerline{ \includegraphics[width=0.48
   \textwidth]{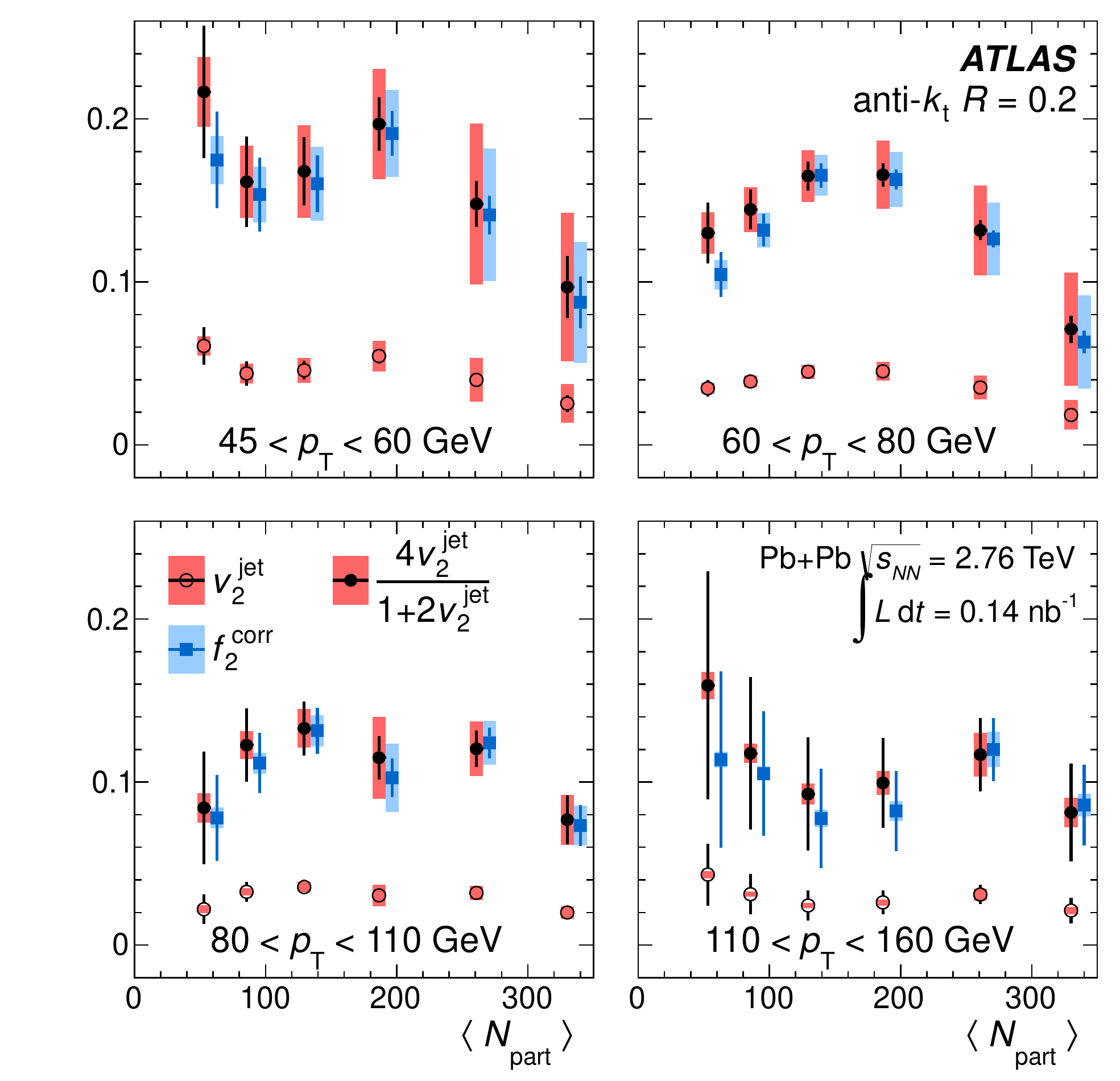} } \caption{   The \ANpart\
   dependence of \vtjet~(\opencircle),
   \ftwocorr~(\textcolor{kAzure}{\fullsquare}) and
   $4\vtjet/(1+2\vtjet)$~(\fullcircle).  All  quantities have
   statistical and systematic uncertainties that are indicated by
   error bars and shaded bands, respectively. The uncertainties for
   all quantities are strongly correlated. The horizontal positions of
   the points have been offset slightly for presentation purposes and
   the width of the error bands indicates the uncertainty on
   \ANpart. } 
\label{fig:jetvtwocent} 
\end{figure}

The centrality dependence of \vtjet\ is shown in
Fig.~\ref{fig:jetvtwocent} as a function of \ANpart\ for different
ranges in \pt. The variation in jet yields with \Dphi\ can also be
characterized by the ratio of jet yields between the most out-of-plane
(\outplane) and most in-plane (\inplane) bins,
\begin{equation} \Rmaxdphi \equiv
\left.\dNdpTdphi\right|_{\mathrm{out}}
/\left.\dNdpTdphi\right|_{\mathrm{in}}\,.
 \end{equation} This quantity is more general than \vtjet\ as it does
not assume a functional form for the \dphi\ dependence of the jet
yields. The nuclear modification factor, \RAA, is a measure of the
effect of quenching on hard scattering rates, and \Rmaxdphi\ can be
interpreted as the ratio of \Dphi-dependent \RAA\ factors, $\Rmaxdphi =
\left.\RAA\right|_{\mathrm{out}}/\left.\RAA\right|_{\mathrm{in}}$
\cite{Adare:2010sp}.  The yields were corrected for \Psit\ resolution
assuming that the \Dphi\ variation is dominated by the $\cos{2\Dphi}$
modulation,
\begin{equation} \dNdpTdphiCorrTall = \dNdpTdphiMeas\left( \dfrac{1 +
2\vtjet\cos{2\Dphi}}{1 + 2\vtjetmeas\cos{2\Dphi}}\right).
 \label{eqn:remodulation}
\end{equation}

The results, expressed in terms of the quantity $\ftwo\equiv 1-
\Rmaxdphi$, show as much as a 20\% variation between the out-of-plane
and in-plane jet yields, but they are reduced slightly from the
maximal difference, evaluated at $\Dphi = \PiOverTwo$ and $\Dphi = 0$,
by the finite bin size used in the measurement. That reduction was
corrected by assuming a $1+2\vtjet \cos2\dphi$ variation of the jet
yields {\em within} the \dphi\ bins containing $\dphi=0$ and
$\PiOverTwo$, and calculating the corresponding yields at those \dphi\
values. From these yields, \ftwocorr\ was calculated analogously to
\ftwo. The magnitude of the correction is typically a few percent. The
\ftwocorr\ values are shown in Fig.~\ref{fig:jetvtwocent}. For a pure
$\cos{2\Dphi}$ modulation of the jet yields, \ftwocorr\ would be given
by $4\vtjet/(1+2\vtjet)$. To test for deviations of the \Dphi\
dependence of the jet yields from a pure $\cos{2\Dphi}$ variation,
$4\vtjet/(1+2\vtjet)$ was calculated using the measured \vtjet\ values
and is shown for each \pT\ and centrality interval in
Fig.~\ref{fig:jetvtwocent}.

Similar variations of \vtjet, \ftwocorr\ and $4\vtjet/(1+2\vtjet)$
with \ANpart\ are seen in the 60\emdash80~\GeV\ range, which has the
best statistical precision.  A reduction in \ftwocorr\ and \vtjet\ in both
the most central and peripheral collisions is not surprising; for very
central collisions, the anisotropy of the initial state is small and
the possible \Dphi\ variation of path lengths in the medium is
limited. Although the anisotropy is greater in peripheral collisions,
there is little suppression in the jet yields
\cite{jetsupp}. Therefore large variations in jet yield as a function
of \Dphi\ would be unexpected. The \ftwocorr\ and
$4\vtjet/(1+2\vtjet)$ values are generally in agreement within
uncertainties, indicating an azimuthal dependence of relative
suppression when measured with respect to the elliptic event plane that is
dominated by second-harmonic modulation.

This \papertype\ has presented results of ATLAS measurements of the
variation of \RTwo\ anti-\kt\ jet yields in $\sqrtsnn =2.76$~\TeV\
\PbPb\ collisions as a function of \Dphi, the relative azimuthal angle of the
jet with respect to the elliptic event plane. A significant \Dphi\
variation in the jet yield is observed for all centrality intervals
and in all \pTjet\ ranges except for the 160\emdash210~\GeV\ \pTjet\
interval where the statistical uncertainties are large. The observed
azimuthal variation of jet yields amounts to a reduction of
10\emdash20\% in the jet yields between in-plane and out-of-plane
directions. These results establish a relationship between jet
suppression and the initial nuclear geometry that should constrain
models of the path-length dependence of the quenching mechanism. 
\par

\par
We thank CERN for the very successful operation of the LHC, as well as the
support staff from our institutions without whom ATLAS could not be
operated efficiently.

We acknowledge the support of ANPCyT, Argentina; YerPhI, Armenia; ARC,
Australia; BMWF and FWF, Austria; ANAS, Azerbaijan; SSTC, Belarus; CNPq and FAPESP,
Brazil; NSERC, NRC and CFI, Canada; CERN; CONICYT, Chile; CAS, MOST and NSFC,
China; COLCIENCIAS, Colombia; MSMT CR, MPO CR and VSC CR, Czech Republic;
DNRF, DNSRC and Lundbeck Foundation, Denmark; EPLANET, ERC and NSRF, European Union;
IN2P3-CNRS, CEA-DSM/IRFU, France; GNSF, Georgia; BMBF, DFG, HGF, MPG and AvH
Foundation, Germany; GSRT and NSRF, Greece; ISF, MINERVA, GIF, DIP and Benoziyo Center,
Israel; INFN, Italy; MEXT and JSPS, Japan; CNRST, Morocco; FOM and NWO,
Netherlands; BRF and RCN, Norway; MNiSW, Poland; GRICES and FCT, Portugal; MERYS
(MECTS), Romania; MES of Russia and ROSATOM, Russian Federation; JINR; MSTD,
Serbia; MSSR, Slovakia; ARRS and MVZT, Slovenia; DST/NRF, South Africa;
MICINN, Spain; SRC and Wallenberg Foundation, Sweden; SER, SNSF and Cantons of
Bern and Geneva, Switzerland; NSC, Taiwan; TAEK, Turkey; STFC, the Royal
Society and Leverhulme Trust, United Kingdom; DOE and NSF, United States of
America.

The crucial computing support from all WLCG partners is acknowledged
gratefully, in particular from CERN and the ATLAS Tier-1 facilities at
TRIUMF (Canada), NDGF (Denmark, Norway, Sweden), CC-IN2P3 (France),
KIT/GridKA (Germany), INFN-CNAF (Italy), NL-T1 (Netherlands), PIC (Spain),
ASGC (Taiwan), RAL (UK) and BNL (USA) and in the Tier-2 facilities
worldwide. 
\bibliography{AzimDep}

\begin{thebibliography}{32}%
\makeatletter
\providecommand \@ifxundefined [1]{%
 \@ifx{#1\undefined}
}%
\providecommand \@ifnum [1]{%
 \ifnum #1\expandafter \@firstoftwo
 \else \expandafter \@secondoftwo
 \fi
}%
\providecommand \@ifx [1]{%
 \ifx #1\expandafter \@firstoftwo
 \else \expandafter \@secondoftwo
 \fi
}%
\providecommand \natexlab [1]{#1}%
\providecommand \enquote  [1]{``#1''}%
\providecommand \bibnamefont  [1]{#1}%
\providecommand \bibfnamefont [1]{#1}%
\providecommand \citenamefont [1]{#1}%
\providecommand \href@noop [0]{\@secondoftwo}%
\providecommand \href [0]{\begingroup \@sanitize@url \@href}%
\providecommand \@href[1]{\@@startlink{#1}\@@href}%
\providecommand \@@href[1]{\endgroup#1\@@endlink}%
\providecommand \@sanitize@url [0]{\catcode `\\12\catcode `\$12\catcode
  `\&12\catcode `\#12\catcode `\^12\catcode `\_12\catcode `\%12\relax}%
\providecommand \@@startlink[1]{}%
\providecommand \@@endlink[0]{}%
\providecommand \url  [0]{\begingroup\@sanitize@url \@url }%
\providecommand \@url [1]{\endgroup\@href {#1}{\urlprefix }}%
\providecommand \urlprefix  [0]{URL }%
\providecommand \Eprint [0]{\href }%
\providecommand \doibase [0]{http://dx.doi.org/}%
\providecommand \selectlanguage [0]{\@gobble}%
\providecommand \bibinfo  [0]{\@secondoftwo}%
\providecommand \bibfield  [0]{\@secondoftwo}%
\providecommand \translation [1]{[#1]}%
\providecommand \BibitemOpen [0]{}%
\providecommand \bibitemStop [0]{}%
\providecommand \bibitemNoStop [0]{.\EOS\space}%
\providecommand \EOS [0]{\spacefactor3000\relax}%
\providecommand \BibitemShut  [1]{\csname bibitem#1\endcsname}%
\let\auto@bib@innerbib\@empty
\bibitem [{\citenamefont {{ATLAS
  Collaboration}}(2010{\natexlab{a}})}]{Aad:2010bu}%
  \BibitemOpen
  \bibfield  {author} {\bibinfo {author} {\bibnamefont {{ATLAS
  Collaboration}}},\ }\href {\doibase 10.1103/PhysRevLett.105.252303}
  {\bibfield  {journal} {\bibinfo  {journal} {Phys.~Rev.~Lett.}\ }\textbf
  {\bibinfo {volume} {105}},\ \bibinfo {pages} {252303} (\bibinfo {year}
  {2010}{\natexlab{a}})}\BibitemShut {NoStop}%
\bibitem [{\citenamefont {{CMS Collaboration}}(2011)}]{Chatrchyan:2011sx}%
  \BibitemOpen
  \bibfield  {author} {\bibinfo {author} {\bibnamefont {{CMS Collaboration}}},\
  }\href {\doibase 10.1103/PhysRevC.84.024906} {\bibfield  {journal} {\bibinfo
  {journal} {Phys.~Rev. C}\ }\textbf {\bibinfo {volume} {84}},\ \bibinfo
  {pages} {024906} (\bibinfo {year} {2011})}\BibitemShut {NoStop}%
\bibitem [{\citenamefont {{ATLAS
  Collaboration}}(2013{\natexlab{a}})}]{jetsupp}%
  \BibitemOpen
  \bibfield  {author} {\bibinfo {author} {\bibnamefont {{ATLAS
  Collaboration}}},\ }\href {\doibase 10.1016/j.physletb.2013.01.024}
  {\bibfield  {journal} {\bibinfo  {journal} {Phys.~Lett. B}\ }\textbf
  {\bibinfo {volume} {719}},\ \bibinfo {pages} {220} (\bibinfo {year}
  {2013}{\natexlab{a}})}\BibitemShut {NoStop}%
\bibitem [{\citenamefont {Baier}\ \emph {et~al.}(1998)\citenamefont {Baier},
  \citenamefont {Dokshitzer}, \citenamefont {Mueller},\ and\ \citenamefont
  {Schiff}}]{Baier:1998yf}%
  \BibitemOpen
  \bibfield  {author} {\bibinfo {author} {\bibfnamefont {R.}~\bibnamefont
  {Baier}}, \bibinfo {author} {\bibfnamefont {Y.~L.}\ \bibnamefont
  {Dokshitzer}}, \bibinfo {author} {\bibfnamefont {A.~H.}\ \bibnamefont
  {Mueller}}, \ and\ \bibinfo {author} {\bibfnamefont {D.}~\bibnamefont
  {Schiff}},\ }\href {\doibase 10.1103/PhysRevC.58.1706} {\bibfield  {journal}
  {\bibinfo  {journal} {Phys.~Rev. C}\ }\textbf {\bibinfo {volume} {58}},\
  \bibinfo {pages} {1706} (\bibinfo {year} {1998})}\BibitemShut {NoStop}%
\bibitem [{\citenamefont {Gubser}\ \emph {et~al.}(2008)\citenamefont {Gubser},
  \citenamefont {Gulotta}, \citenamefont {Pufu},\ and\ \citenamefont
  {Rocha}}]{Gubser:2008as}%
  \BibitemOpen
  \bibfield  {author} {\bibinfo {author} {\bibfnamefont {S.~S.}\ \bibnamefont
  {Gubser}}, \bibinfo {author} {\bibfnamefont {D.~R.}\ \bibnamefont {Gulotta}},
  \bibinfo {author} {\bibfnamefont {S.~S.}\ \bibnamefont {Pufu}}, \ and\
  \bibinfo {author} {\bibfnamefont {F.~D.}\ \bibnamefont {Rocha}},\ }\href
  {\doibase 10.1088/1126-6708/2008/10/052} {\bibfield  {journal} {\bibinfo
  {journal} {J.~High~Energy~Phys.}\ }\bibinfo {series} {10},\ \bibinfo {pages}
  {052} (\bibinfo {year} {2008})}\BibitemShut {NoStop}%
\bibitem [{\citenamefont {Marquet}\ and\ \citenamefont
  {Renk}(2010)}]{Marquet:2009eq}%
  \BibitemOpen
  \bibfield  {author} {\bibinfo {author} {\bibfnamefont {C.}~\bibnamefont
  {Marquet}}\ and\ \bibinfo {author} {\bibfnamefont {T.}~\bibnamefont {Renk}},\
  }\href {\doibase 10.1016/j.physletb.2010.01.076} {\bibfield  {journal}
  {\bibinfo  {journal} {Phys.~Lett. B}\ }\textbf {\bibinfo {volume} {685}},\
  \bibinfo {pages} {270} (\bibinfo {year} {2010})}\BibitemShut {NoStop}%
\bibitem [{\citenamefont {Gyulassy}\ \emph {et~al.}(2001)\citenamefont
  {Gyulassy}, \citenamefont {Vitev},\ and\ \citenamefont
  {Wang}}]{Gyulassy:2000gk}%
  \BibitemOpen
  \bibfield  {author} {\bibinfo {author} {\bibfnamefont {M.}~\bibnamefont
  {Gyulassy}}, \bibinfo {author} {\bibfnamefont {I.}~\bibnamefont {Vitev}}, \
  and\ \bibinfo {author} {\bibfnamefont {X.~N.}\ \bibnamefont {Wang}},\ }\href
  {\doibase 10.1103/PhysRevLett.86.2537} {\bibfield  {journal} {\bibinfo
  {journal} {Phys.~Rev.~Lett.}\ }\textbf {\bibinfo {volume} {86}},\ \bibinfo
  {pages} {2537} (\bibinfo {year} {2001})}\BibitemShut {NoStop}%
\bibitem [{\citenamefont {Zhang}\ and\ \citenamefont
  {Liao}(2012)}]{Zhang:2012ha}%
  \BibitemOpen
  \bibfield  {author} {\bibinfo {author} {\bibfnamefont {X.}~\bibnamefont
  {Zhang}}\ and\ \bibinfo {author} {\bibfnamefont {J.}~\bibnamefont {Liao}},\
  }\href@noop {} {\  (\bibinfo {year} {2012})},\ \Eprint
  {http://arxiv.org/abs/1210.1245} {arXiv:1210.1245} \BibitemShut {NoStop}%
\bibitem [{Note1()}]{Note1}%
  \BibitemOpen
  \bibinfo {note} {ATLAS uses a right-handed coordinate system with its origin
  at the nominal interaction point (IP) in the center of the detector and the
  $z$-axis along the beam pipe. The $x$-axis points from the IP to the center
  of the LHC ring, and the $y$-axis points upward. Cylindrical coordinates
  $(r,\phi )$ are used in the transverse plane, $\phi $ being the azimuthal
  angle around the beam pipe. The pseudorapidity is defined in terms of the
  polar angle $\theta $ as $\protect \ensuremath {\etaa }=-\protect \qopname
  \relax o{ln}\protect \qopname \relax o{tan}(\theta /2)$.}\BibitemShut {Stop}%
\bibitem [{\citenamefont {Ollitrault}(1992)}]{Ollitrault:1992bk}%
  \BibitemOpen
  \bibfield  {author} {\bibinfo {author} {\bibfnamefont {J.-Y.}\ \bibnamefont
  {Ollitrault}},\ }\href {\doibase 10.1103/PhysRevD.46.229} {\bibfield
  {journal} {\bibinfo  {journal} {Phys.~Rev. D}\ }\textbf {\bibinfo {volume}
  {46}},\ \bibinfo {pages} {229} (\bibinfo {year} {1992})}\BibitemShut
  {NoStop}%
\bibitem [{\citenamefont {{ALICE Collaboration}}(2011)}]{ALICE:2011ab}%
  \BibitemOpen
  \bibfield  {author} {\bibinfo {author} {\bibnamefont {{ALICE
  Collaboration}}},\ }\href {\doibase 10.1103/PhysRevLett.107.032301}
  {\bibfield  {journal} {\bibinfo  {journal} {Phys.~Rev.~Lett.}\ }\textbf
  {\bibinfo {volume} {107}},\ \bibinfo {pages} {032301} (\bibinfo {year}
  {2011})}\BibitemShut {NoStop}%
\bibitem [{\citenamefont {{ATLAS
  Collaboration}}(2012{\natexlab{a}})}]{ATLAS:2012at}%
  \BibitemOpen
  \bibfield  {author} {\bibinfo {author} {\bibnamefont {{ATLAS
  Collaboration}}},\ }\href {\doibase 10.1103/PhysRevC.86.014907} {\bibfield
  {journal} {\bibinfo  {journal} {Phys.~Rev. C}\ }\textbf {\bibinfo {volume}
  {86}},\ \bibinfo {pages} {014907} (\bibinfo {year}
  {2012}{\natexlab{a}})}\BibitemShut {NoStop}%
\bibitem [{\citenamefont {{CMS
  Collaboration}}(2012{\natexlab{a}})}]{Chatrchyan:2012wg}%
  \BibitemOpen
  \bibfield  {author} {\bibinfo {author} {\bibnamefont {{CMS Collaboration}}},\
  }\href {\doibase 10.1140/epjc/s10052-012-2012-3} {\bibfield  {journal}
  {\bibinfo  {journal} {Eur.~Phys.~J. C}\ }\textbf {\bibinfo {volume} {72}},\
  \bibinfo {pages} {2012} (\bibinfo {year} {2012}{\natexlab{a}})}\BibitemShut
  {NoStop}%
\bibitem [{\citenamefont {{STAR Collaboration}}(2004)}]{Adams:2004wz}%
  \BibitemOpen
  \bibfield  {author} {\bibinfo {author} {\bibnamefont {{STAR
  Collaboration}}},\ }\href {\doibase 10.1103/PhysRevLett.93.252301} {\bibfield
   {journal} {\bibinfo  {journal} {Phys.~Rev.~Lett.}\ }\textbf {\bibinfo
  {volume} {93}},\ \bibinfo {pages} {252301} (\bibinfo {year}
  {2004})}\BibitemShut {NoStop}%
\bibitem [{\citenamefont {{PHENIX Collaboration}}(2007)}]{Adler:2006bw}%
  \BibitemOpen
  \bibfield  {author} {\bibinfo {author} {\bibnamefont {{PHENIX
  Collaboration}}},\ }\href {\doibase 10.1103/PhysRevC.76.034904} {\bibfield
  {journal} {\bibinfo  {journal} {Phys.~Rev. C}\ }\textbf {\bibinfo {volume}
  {76}},\ \bibinfo {pages} {034904} (\bibinfo {year} {2007})}\BibitemShut
  {NoStop}%
\bibitem [{\citenamefont {{PHENIX Collaboration}}(2010)}]{Adare:2010sp}%
  \BibitemOpen
  \bibfield  {author} {\bibinfo {author} {\bibnamefont {{PHENIX
  Collaboration}}},\ }\href {\doibase 10.1103/PhysRevLett.105.142301}
  {\bibfield  {journal} {\bibinfo  {journal} {Phys.~Rev.~Lett.}\ }\textbf
  {\bibinfo {volume} {105}},\ \bibinfo {pages} {142301} (\bibinfo {year}
  {2010})}\BibitemShut {NoStop}%
\bibitem [{\citenamefont {{ATLAS
  Collaboration}}(2012{\natexlab{b}})}]{ATLAS:2011ah}%
  \BibitemOpen
  \bibfield  {author} {\bibinfo {author} {\bibnamefont {{ATLAS
  Collaboration}}},\ }\href {\doibase 10.1016/j.physletb.2011.12.056}
  {\bibfield  {journal} {\bibinfo  {journal} {Phys.~Lett. B}\ }\textbf
  {\bibinfo {volume} {707}},\ \bibinfo {pages} {330} (\bibinfo {year}
  {2012}{\natexlab{b}})}\BibitemShut {NoStop}%
\bibitem [{\citenamefont {{CMS
  Collaboration}}(2012{\natexlab{b}})}]{Chatrchyan:2012xq}%
  \BibitemOpen
  \bibfield  {author} {\bibinfo {author} {\bibnamefont {{CMS Collaboration}}},\
  }\href {\doibase 10.1103/PhysRevLett.109.022301} {\bibfield  {journal}
  {\bibinfo  {journal} {Phys.~Rev.~Lett.}\ }\textbf {\bibinfo {volume} {109}},\
  \bibinfo {pages} {022301} (\bibinfo {year} {2012}{\natexlab{b}})}\BibitemShut
  {NoStop}%
\bibitem [{\citenamefont {{ALICE Collaboration}}(2013)}]{Abelev:2012di}%
  \BibitemOpen
  \bibfield  {author} {\bibinfo {author} {\bibnamefont {{ALICE
  Collaboration}}},\ }\href {\doibase 10.1016/j.physletb.2012.12.066}
  {\bibfield  {journal} {\bibinfo  {journal} {Phys.~Lett. B}\ }\textbf
  {\bibinfo {volume} {719}},\ \bibinfo {pages} {18} (\bibinfo {year}
  {2013})}\BibitemShut {NoStop}%
\bibitem [{\citenamefont {Cacciari}\ \emph {et~al.}(2008)\citenamefont
  {Cacciari}, \citenamefont {Salam},\ and\ \citenamefont
  {Soyez}}]{Cacciari:2008gp}%
  \BibitemOpen
  \bibfield  {author} {\bibinfo {author} {\bibfnamefont {M.}~\bibnamefont
  {Cacciari}}, \bibinfo {author} {\bibfnamefont {G.~P.}\ \bibnamefont {Salam}},
  \ and\ \bibinfo {author} {\bibfnamefont {G.}~\bibnamefont {Soyez}},\ }\href
  {\doibase 10.1088/1126-6708/2008/04/063} {\bibfield  {journal} {\bibinfo
  {journal} {J.~High~Energy~Phys.}\ }\bibinfo {series} {04},\ \bibinfo {pages}
  {063} (\bibinfo {year} {2008})}\BibitemShut {NoStop}%
\bibitem [{\citenamefont {{ATLAS Collaboration}}(2008)}]{Aad:2008zzm}%
  \BibitemOpen
  \bibfield  {author} {\bibinfo {author} {\bibnamefont {{ATLAS
  Collaboration}}},\ }\href {\doibase 10.1088/1748-0221/3/08/S08003} {\bibfield
   {journal} {\bibinfo  {journal} {JINST}\ }\textbf {\bibinfo {volume} {3}},\
  \bibinfo {pages} {S08003} (\bibinfo {year} {2008})}\BibitemShut {NoStop}%
\bibitem [{\citenamefont {{ATLAS
  Collaboration}}(2010{\natexlab{b}})}]{Aad:2010bx}%
  \BibitemOpen
  \bibfield  {author} {\bibinfo {author} {\bibnamefont {{ATLAS
  Collaboration}}},\ }\href {\doibase 10.1140/epjc/s10052-010-1366-7}
  {\bibfield  {journal} {\bibinfo  {journal} {Eur.~Phys.~J. C}\ }\textbf
  {\bibinfo {volume} {70}},\ \bibinfo {pages} {787} (\bibinfo {year}
  {2010}{\natexlab{b}})}\BibitemShut {NoStop}%
\bibitem [{\citenamefont {Poskanzer}\ and\ \citenamefont
  {Voloshin}(1998)}]{Poskanzer:1998yz}%
  \BibitemOpen
  \bibfield  {author} {\bibinfo {author} {\bibfnamefont {A.~M.}\ \bibnamefont
  {Poskanzer}}\ and\ \bibinfo {author} {\bibfnamefont {S.~A.}\ \bibnamefont
  {Voloshin}},\ }\href {\doibase 10.1103/PhysRevC.58.1671} {\bibfield
  {journal} {\bibinfo  {journal} {Phys.~Rev. C}\ }\textbf {\bibinfo {volume}
  {58}},\ \bibinfo {pages} {1671} (\bibinfo {year} {1998})}\BibitemShut
  {NoStop}%
\bibitem [{\citenamefont {Miller}\ \emph {et~al.}(2007)\citenamefont {Miller},
  \citenamefont {Reygers}, \citenamefont {Sanders},\ and\ \citenamefont
  {Steinberg}}]{Miller:2007ri}%
  \BibitemOpen
  \bibfield  {author} {\bibinfo {author} {\bibfnamefont {M.~L.}\ \bibnamefont
  {Miller}}, \bibinfo {author} {\bibfnamefont {K.}~\bibnamefont {Reygers}},
  \bibinfo {author} {\bibfnamefont {S.~J.}\ \bibnamefont {Sanders}}, \ and\
  \bibinfo {author} {\bibfnamefont {P.}~\bibnamefont {Steinberg}},\ }\href
  {\doibase 10.1146/annurev.nucl.57.090506.123020} {\bibfield  {journal}
  {\bibinfo  {journal} {Ann.~Rev.~Nucl.~Part.~Sci.}\ }\textbf {\bibinfo
  {volume} {57}},\ \bibinfo {pages} {205} (\bibinfo {year} {2007})}\BibitemShut
  {NoStop}%
\bibitem [{\citenamefont {Alver}\ \emph {et~al.}(2008)\citenamefont {Alver},
  \citenamefont {Baker}, \citenamefont {Loizides},\ and\ \citenamefont
  {Steinberg}}]{Alver:2008aq}%
  \BibitemOpen
  \bibfield  {author} {\bibinfo {author} {\bibfnamefont {B.}~\bibnamefont
  {Alver}}, \bibinfo {author} {\bibfnamefont {M.}~\bibnamefont {Baker}},
  \bibinfo {author} {\bibfnamefont {C.}~\bibnamefont {Loizides}}, \ and\
  \bibinfo {author} {\bibfnamefont {P.}~\bibnamefont {Steinberg}},\ }\href@noop
  {} {\  (\bibinfo {year} {2008})},\ \Eprint {http://arxiv.org/abs/0805.4411}
  {arXiv:0805.4411} \BibitemShut {NoStop}%
\bibitem [{\citenamefont {{ATLAS
  Collaboration}}(2013{\natexlab{b}})}]{Aad:2011he}%
  \BibitemOpen
  \bibfield  {author} {\bibinfo {author} {\bibnamefont {{ATLAS
  Collaboration}}},\ }\href {\doibase 10.1140/epjc/s10052-013-2304-2}
  {\bibfield  {journal} {\bibinfo  {journal} {Eur.~Phys.~J. C}\ }\textbf
  {\bibinfo {volume} {73}},\ \bibinfo {pages} {2304} (\bibinfo {year}
  {2013}{\natexlab{b}})}\BibitemShut {NoStop}%
\bibitem [{\citenamefont {{ATLAS Collaboration}}(2009)}]{Aad:2009wy}%
  \BibitemOpen
  \bibfield  {author} {\bibinfo {author} {\bibnamefont {{ATLAS
  Collaboration}}},\ }\href@noop {} {\  (\bibinfo {year} {2009})},\ \Eprint
  {http://arxiv.org/abs/0901.0512} {arXiv:0901.0512} \BibitemShut {NoStop}%
\bibitem [{Note2()}]{Note2}%
  \BibitemOpen
  \bibinfo {note} {This differs from the requirement used in Ref.~\cite
  {jetsupp}, where the minimum \protect \ensuremath {p_{\protect \mathrm {T}}}\
  was 7~\protect \textrm {Ge\kern -0.1em V}\ for both track jets and
  clusters.}\BibitemShut {Stop}%
\bibitem [{\citenamefont {Agostinelli}\ \emph {et~al.}(2003)\citenamefont
  {Agostinelli} \emph {et~al.}}]{Agostinelli:2002hh}%
  \BibitemOpen
  \bibfield  {author} {\bibinfo {author} {\bibfnamefont {S.}~\bibnamefont
  {Agostinelli}} \emph {et~al.} (\bibinfo {collaboration} {GEANT4}),\ }\href
  {\doibase 10.1016/S0168-9002(03)01368-8} {\bibfield  {journal} {\bibinfo
  {journal} {Nucl.~Instrum.~Meth. A}\ }\textbf {\bibinfo {volume} {506}},\
  \bibinfo {pages} {250} (\bibinfo {year} {2003})}\BibitemShut {NoStop}%
\bibitem [{\citenamefont {{ATLAS
  Collaboration}}(2010{\natexlab{c}})}]{atlassim}%
  \BibitemOpen
  \bibfield  {author} {\bibinfo {author} {\bibnamefont {{ATLAS
  Collaboration}}},\ }\href@noop {} {\bibfield  {journal} {\bibinfo  {journal}
  {Eur.~Phys.~J. C}\ }\textbf {\bibinfo {volume} {70}},\ \bibinfo {pages} {823}
  (\bibinfo {year} {2010}{\natexlab{c}})}\BibitemShut {NoStop}%
\bibitem [{\citenamefont {Sjostrand}\ \emph {et~al.}(2006)\citenamefont
  {Sjostrand}, \citenamefont {Mrenna},\ and\ \citenamefont
  {Skands}}]{Sjostrand:2006za}%
  \BibitemOpen
  \bibfield  {author} {\bibinfo {author} {\bibfnamefont {T.}~\bibnamefont
  {Sjostrand}}, \bibinfo {author} {\bibfnamefont {S.}~\bibnamefont {Mrenna}}, \
  and\ \bibinfo {author} {\bibfnamefont {P.}~\bibnamefont {Skands}},\ }\href
  {\doibase 10.1088/1126-6708/2006/05/026} {\bibfield  {journal} {\bibinfo
  {journal} {J.~High~Energy~Phys.}\ }\textbf {\bibinfo {volume} {05}},\
  \bibinfo {pages} {026} (\bibinfo {year} {2006})}\BibitemShut {NoStop}%
\bibitem [{\citenamefont {{ATLAS Collaboration}}(2011)}]{mctunes}%
  \BibitemOpen
  \bibfield  {author} {\bibinfo {author} {\bibnamefont {{ATLAS
  Collaboration}}},\ }\href {http://cdsweb.cern.ch/record/1363300} {\
  (\bibinfo {year} {2011})},\ \Eprint {{}/ATL-PHYS-PUB-2011-009}
  {ATL-PHYS-PUB-2011-009} \BibitemShut {NoStop}%
\end{thebibliography}%
\bibliographystyle{apsrev4-1}

\onecolumngrid
\section*{The ATLAS Collaboration}
\label{app:ATLASColl}
\begin{flushleft}

\bigskip

G.~Aad$^{\rm 48}$,
T.~Abajyan$^{\rm 21}$,
B.~Abbott$^{\rm 112}$,
J.~Abdallah$^{\rm 12}$,
S.~Abdel~Khalek$^{\rm 116}$,
O.~Abdinov$^{\rm 11}$,
R.~Aben$^{\rm 106}$,
B.~Abi$^{\rm 113}$,
M.~Abolins$^{\rm 89}$,
O.S.~AbouZeid$^{\rm 159}$
% ...
\bigskip
\\
% ...
$^{11}$ Institute of Physics, Azerbaijan Academy of Sciences, Baku, Azerbaijan\\
$^{12}$ Institut de F{\'\i}sica d'Altes Energies and Departament de F{\'\i}sica de la Universitat Aut{\`o}noma de Barcelona, Barcelona, Spain\\
% ...
$^{21}$ Physikalisches Institut, University of Bonn, Bonn, Germany\\
% ...
$^{48}$ Fakult{\"a}t f{\"u}r Mathematik und Physik, Albert-Ludwigs-Universit{\"a}t, Freiburg, Germany\\
% ...
$^{89}$ Department of Physics and Astronomy, Michigan State University, East Lansing MI, United States of America\\
% ...
$^{106}$ Nikhef National Institute for Subatomic Physics and University of Amsterdam, Amsterdam, Netherlands\\
% ...
$^{112}$ Homer L. Dodge Department of Physics and Astronomy, University of Oklahoma, Norman OK, United States of America\\
$^{113}$ Department of Physics, Oklahoma State University, Stillwater OK, United States of America\\
% ...
$^{116}$ LAL, Universit{\'e} Paris-Sud and CNRS/IN2P3, Orsay, France\\
% ...
$^{159}$ Department of Physics, University of Toronto, Toronto ON, Canada\\
% ...
\end{flushleft}


\end{document}